\documentclass[a4paper,11pt]{article}
\usepackage{graphicx}
\usepackage[authoryear]{natbib} 
\usepackage[affil-it]{authblk}
\usepackage[colorlinks=true,linkcolor=blue,urlcolor=blue,citecolor=blue]{hyperref}

\title{Around the Pleiades}
\author{Guillermo Abramson\footnote{E-mail: abramson@cab.cnea.gov.ar.}}
\affil{Centro At\'{o}mico Bariloche, CONICET and Instituto Balseiro\\R8402AGP Bariloche, Argentina}

\begin{document}
\maketitle

\begin{abstract}
We present a calculation of the distance to the Pleiades star cluster based on data from Gaia DR2. We show that Gaia finally settles the discrepancy between the values derived from Hipparcos and other distance determinations. The technical level of the presentation is adequate for the interested layperson.
\end{abstract}

How far are the stars? One would say that astronomers should know. It is, however, something very difficult to find out. In fact, the efforts to measure the distance to the stars have run for centuries of development of astronmy since classical antiquity. The stars are actually so far away that no method succeeded until the Industrial Revolution provided exquisitely precise instruments of observation and measurement in the 19th century. With them we could measure the distance to the stars using the same phenomenon that allows us to estimate distances in daily life, from threading a needle to catching a ball in mid-air: our eyes see the world from slightly different perspectives, and our brain porcesses this difference to build a three-dimensional image of our environment. 

\begin{figure}
\centering
\includegraphics[width=0.6\textwidth]{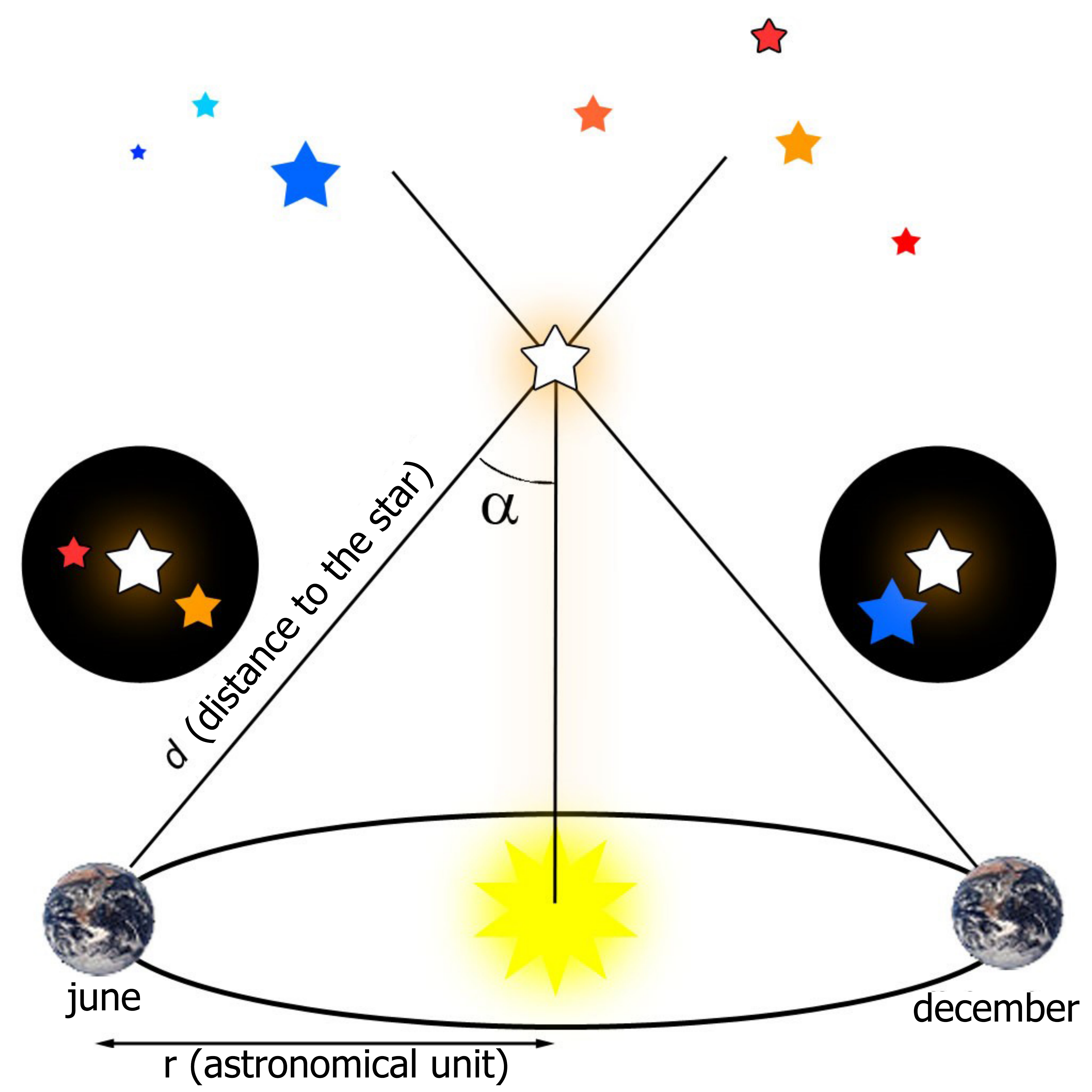}
\caption{Stellar parallax is the apparent variation of the position of a star when the Earth moves along its orbit. The angle $\alpha$ (hugely exagerated in the illustration), measured with the Earth at opposite sites of its orbit, gives the distance by an elementary trigonometric calculation.}
\label{parallax}
\end{figure}

In astronomy, the method consists in observing the apparent shift of the position of a star while the Earth moves along its orbit (Fig.~\ref{parallax}). This can be accomplished by measuring the angle $\alpha$, called \emph{stellar parallax}. It turns out that, even for the closest stars, this angle is extremely small. It wasn't until 1838 that Friedrich Bessel could measure, beyond controversy, the parallax of a star: that of 61 Cygni, which turned out to be 10 light years from us. Its parallax, a mere third of an arcsecond\footnote{One arcsecond is $1/3600$ of a degree. The Moon (and the Sun) occupy half a degree in the sky (1800 seconds).}, is equivalent to discerning a car from 3000 km away. During the whole century that followed barely a hundred stellar paralaxes could be measured. The massive measurement of hundreds of thousands of stars is a task that had to wait for modern technology \citep{hirshfeld2001,abramson2010}.

Near the end of the 20th century the European Space Agency (ESA) designed a space telescope specifically to measure stellar parallaxes. The satellite, called Hipparcos in honour of the Greek astronomer Hipparchus of Nicaea (2nd century BCE), observed a predefined set of stars during four years. The result was the Hipparcos Catalogue, published in 1993, containing high precision parallaxes of a little more than a $100~\!000$ stars. The precision achieved was about 1~mas (one milliarcsecond, like a hair 20 km away\dots). All of them are within a sphere some 300 light years around the Sun. In a galaxy 100 thousand light years in diameter, there was still a lot to explore.

Even though the Hipparcos mission was extremely successful, several surprises were immediately aparent. The most notable was the distance to me most famous of stellar clusters, the Pleiades (Messier 45). Hipparcos found a distance of 115~pc (parsecs\footnote{One parsec is approximately 3.26 light years.}), rather less than the 130~pc of previous calculations, which were based on their brightness and considerations of stellar physics. 

This was an embarrassing problem. On the one hand, the Pleiades are a nearby cluster, and for this reason very important for astronomy. Open clusters play a crucial role because their stars are of the same age, having formed together from the same interstellar cloud. As such, they are excellent laboratories to test physical models of stellar evolution. But for this it is necessary to know their distance and derive their intrinsic brightness. In turn, these models allow to calculate the distance to farther stars, those that are removed from the reach of direct geometrical methods. They play a keystone role in the calibration of the cosmic distances ladder, which procedes step by step, from the Sun to the nearest stars, then to far stars, and so on and so forth, changing methods along the way until the confines of the universe. Much astronomy, from stellar physics to the structure and evolution of the universe, depends on a good calibration of this distance ladder.

On the other hand a dubious result challenged all the Hipparcos Catalogue. Was there some instrumental or systematic error that had been overlooked? Was there a problem with just the Pleiades, or also with other measurements? Or were the Pleiades really closer, and they just didn't fit into the models of stellar formation and evolution?

It took many years to solve the issue, and it is not completely clear what really happened. It is apparently a matter of calibration of the instrument, due to the intrincate method of observation of the telescope. Instead of looking at a fixed spot in the sky (like any other telescope does), Hipparcos rotated about itself, a common strategy to keep satellites very stable. The optical instrument looked ``sideways,'' recording the stars as circular traces. Closely packed stars---like those in a cluster---gave tight traces in such a way that, despite being independent stars, there measurements were closely correlated. This called for different calibrations at different spatial scales, and resulted in an unexpected source of error for the important and compact open clusters.

\begin{figure}
\centering
\includegraphics[width=0.6\textwidth]{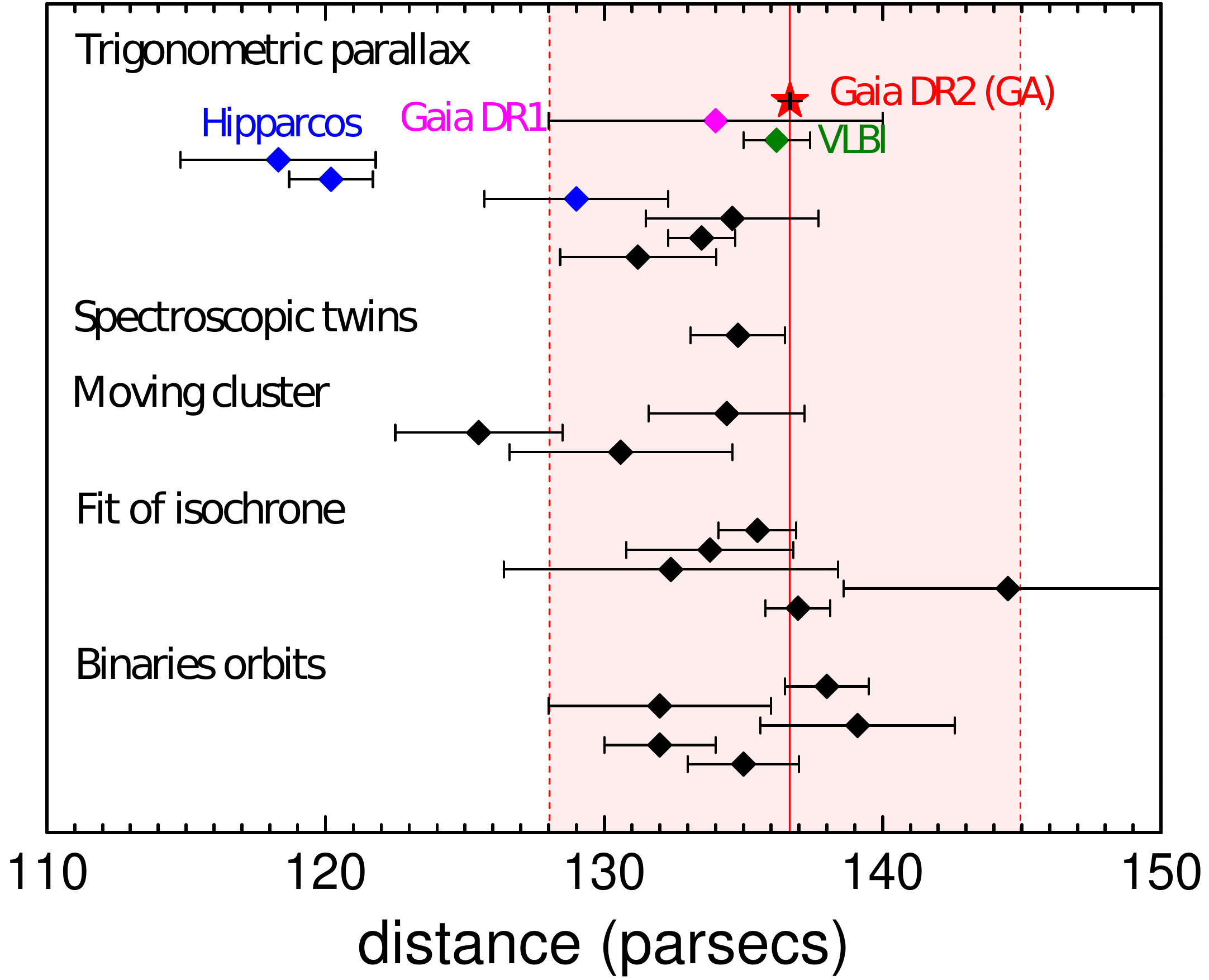}
\caption{Distance to the Pleiades according a variety of methods. Note the anomalous distance measured by Hipparcos and a successive recalibration (blue), the precise result of VLBI (green), the preliminar one given by Gaia DR1 (magenta) and the one presented in this work, based on Gaia DR2 (red). The pink range shows the standard deviation of the sample representative of the stellar distribution in the cluster). Data from the bibliography found in \citep{melis2014}, except the one indicated GA.}
\label{comparacion}
\end{figure}

Additional measurements made with other instruments and methods came to confirm this suspicion. The Pleiades, after all, were where everybody expected and not where Hipparcos said. In 2005 a measurement of 3 stars of the Pleiades done by the Hubble Space Telescope gave a distance of 133.5~pc \citep{soderblom2005}. In 2014 an extremely exact and precise measurement made using radio telescopes all over the world as if they were a single instrument (Very Large Baseline Interferometry, VLBI) gave a result of 136.2~pc \citep{melis2014}. In Fig.~\ref{comparacion} we show some of these measurements, and one can see that those from Hipparcos look anomalously small, even though successive reassessments of the data allowed to correct the initial systematic errors to some extent. 

The successor of Hipparcos is Gaia, also an ESA satellite. Like Hipparcos, it is a telescope of unusual design: it is shaped like a hat, its mirrors are rectangular and it also looks sideways while it spins. Its impressive catalogue Gaia Data Release 2 was published in April this year. It contains the position, parallax and proper motion of more than 1300 million sources up to magnitude 21, reaching even the faraway center of our galaxy. The uncertainty of its parallaxes is around 40~mas up to magnitude 15. Gaia also measured the radial velocity (approaching or receding from us) of more than 7 million stars, brightness of more than 1600 millions with precision of a millimagnitude, temperature of more than 160 millions, classified more than half a million variable stars, observed more than 14 thousand solar system objects and much more. Gaia is still taking data, and there will be third catalogue in 2020 and a final one in 2022 (or later if the mission extends for one more decade). Besides, it has hopefully overcome some of the systematic problems encountered by Hipparcos. What value would Gaia find for the distance to the famous cluster? One of the articles of the preliminar Gaia Data Release 1 \citep{dr1} shows the Pleiades precisely as an example. Using the preliminar values of 164 stars belonging to the cluster they find a distance of 134~pc (see Fig.~\ref{comparacion}), thus confirming the erroneous value of Hipparcos. Using the same method presented there we offer, in the following, a calculation of the distance to the Pleiades based on Gaia DR2. 

\begin{figure}[t]
\centering
\includegraphics[width=0.8\textwidth]{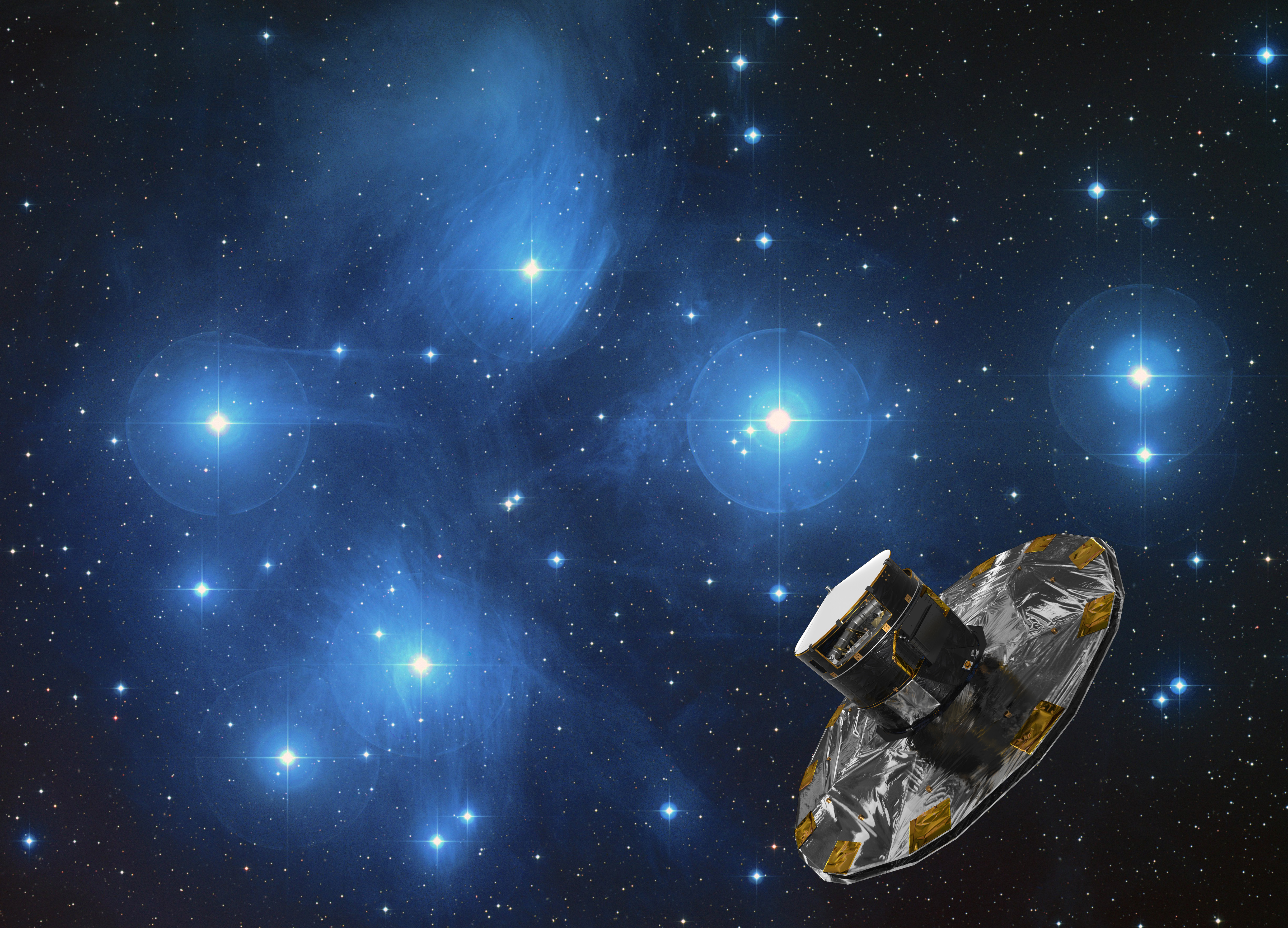}
\caption{Artist representation of Gaia, a space telescope of unusual design, specifically prepared for high precision astrometry and photometry. (Credit: ESA/ATG medialab; background image by NASA/ESA/AURA/Caltech.)}
\label{gaia}
\end{figure}

A download of all the (nearly $700~\!000$!) sources centered in the position of the Pleiades up to some radius gives a cone of observations, with its tip in the solar system and extending indefinitely in space. Somewhere inside that cone lie the Pleiades, as well as many field stars in front of and beyond the cluster. The extraction of the Pleiades from such a large stellar population is, fortunately, quite simple. It is based on the fact that the stars of an open cluster move together and share the same proper motion in the sky, which was also measured by Gaia. According to \citet{dr1} the Pleiades can be identified as having a proper motion around 50~mas per year, in a particular southeastern direction (more details in the Appendix). This leaves us with 1876 stars. Figure~\ref{histogram} shows a histogram of their parallaxes. We can clearly see a peak at about 7~mas, which are 140~pc (around 400 light years), so these are the Pleiades. We can also see some additional ``field stars'': stars with the same movement in the sky but which are much farther away (smaller parallaxes). Selecting those stars with parallaxes between 5 and 9.5 we obtain 1594 stars, 10 times more than those used in \citep{dr1}.

\begin{figure}
\centering
\includegraphics[width=0.7\textwidth]{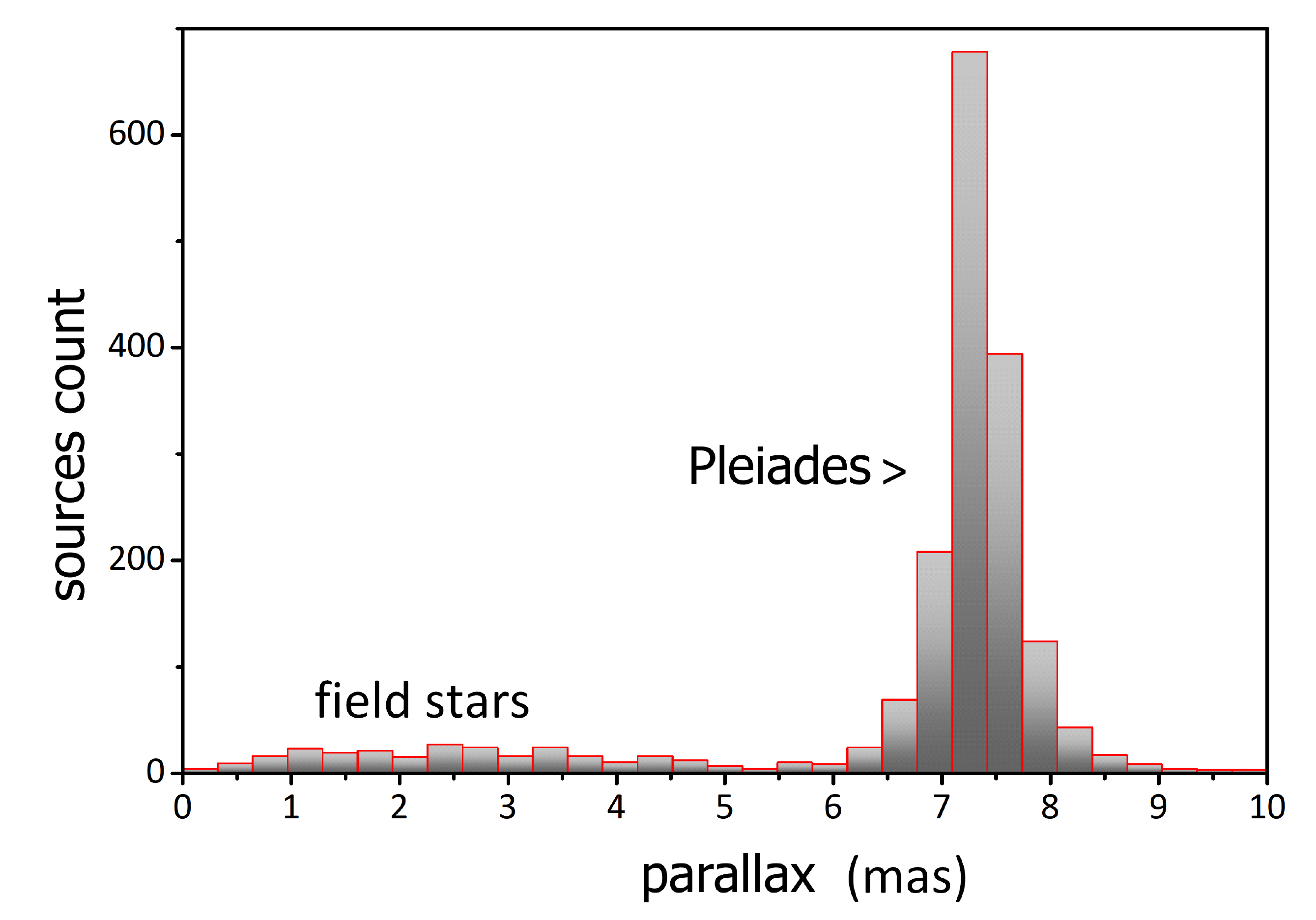}
\caption{Histogram of the parallaxes of the sources of Gaia DR2 in the field of the Pleiades, and with similar proper motion.}
\label{histogram}
\end{figure}

The average parallax of this set is 7.34~mas, with a standard deviation of 0.45~mas. The Pleiades are so close that this deviation is not actually a measurement error, but rather a statistical characterization of the distribution of the stars of the cluster around its center (shown as a  pink shade in Fig.~\ref{comparacion}). But we can do even better than this. As every scientific measurement, each parallax of Gaia is affected by an uncertainty, also reported in the catalog. These errors cover a rather large range of values, from 20~$\mu$as (microarcseconds, $10^{-6}$ seconds of arc) to a couple milliarcseconds. Clearly not all the data are equally confident, and this can be taken into account very precisely in the calculation of the average and its error. A weighted mean\footnote{See detailes in the Appendix. This calculation is an improvement with respect to the initial version of this paper.} of the data gives us a very precise final result of $7.317\pm 0.002$ mas, which corresponds to a distance of $136.67\pm 0.04$~pc  (the error bars are not even visible at the scale of Fig.~\ref{comparacion}) or $445.8\pm 0.1$ light years.

It comes almost as a relief: the Pleiades are where they should, and stellar physics is all right. It is worth noting that 7.3 milliarcseconds is like seeing a hair 3 kilometers away.

\begin{figure}
\includegraphics[width=\textwidth]{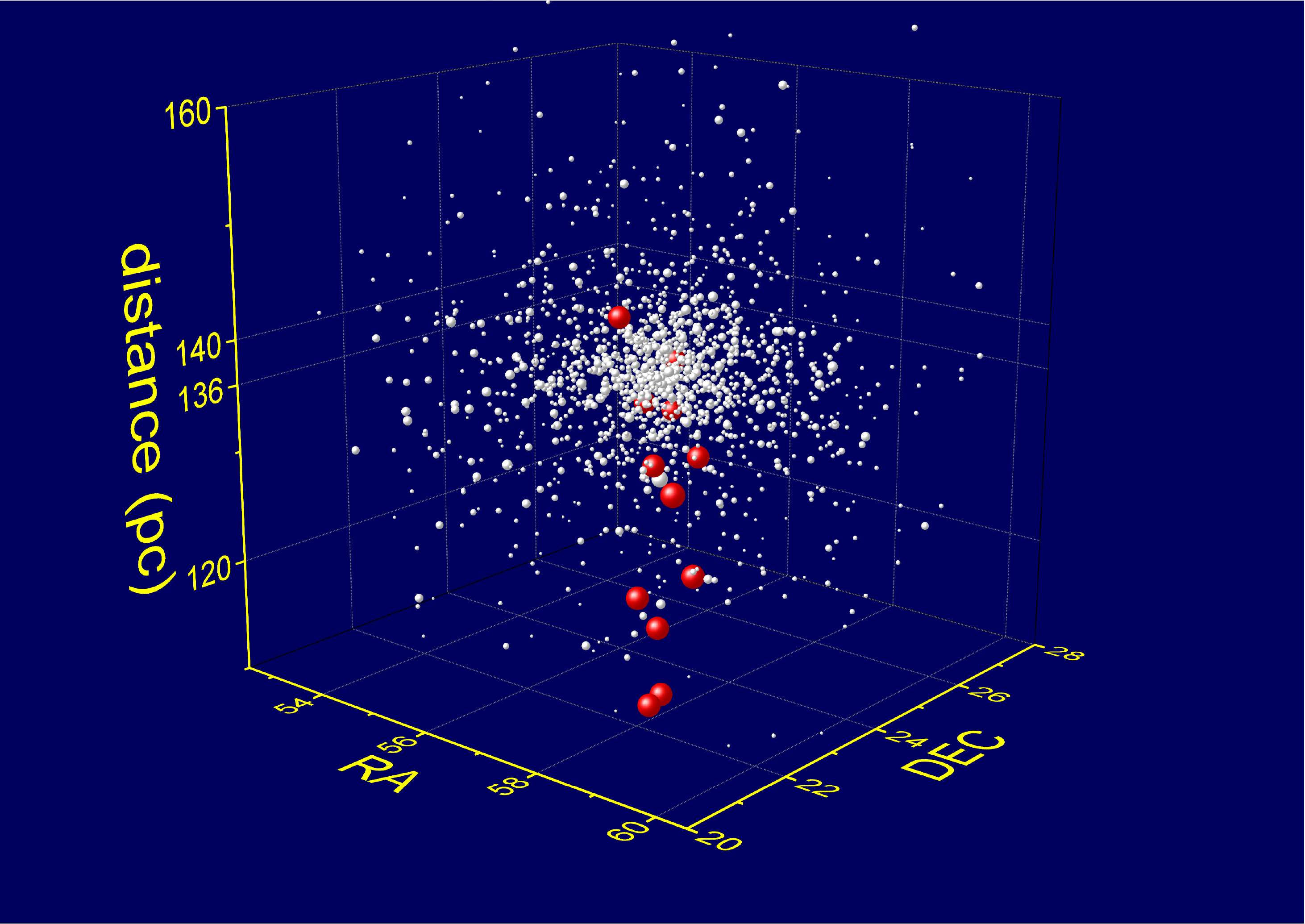}
\caption{Three dimensional distribution of the Pleiades star cluster. The size of the dots corresponds to magnitude. The 12 stars with mag $\le 6$ are highlighted in red, with their size multiplied by a factor 2.}
\label{3dplot}
\end{figure}

\begin{figure}[t]
\includegraphics[width=\textwidth]{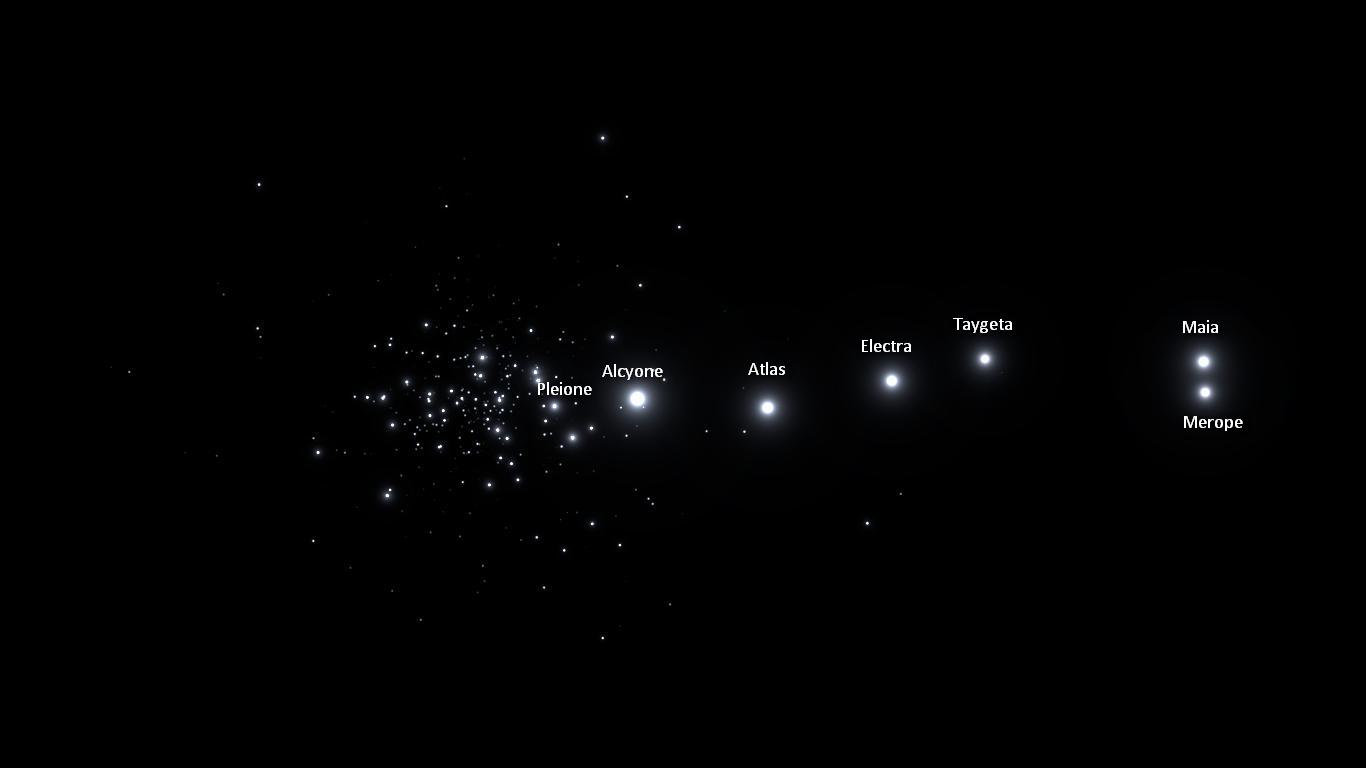}
\caption{Realistic visualization of the Pleiades as seen from an unusual perspective. The solar system lies towards the right of the image. (Done with Celestia.)}
\label{sideways}
\end{figure}

\begin{figure}
\includegraphics[width=\textwidth]{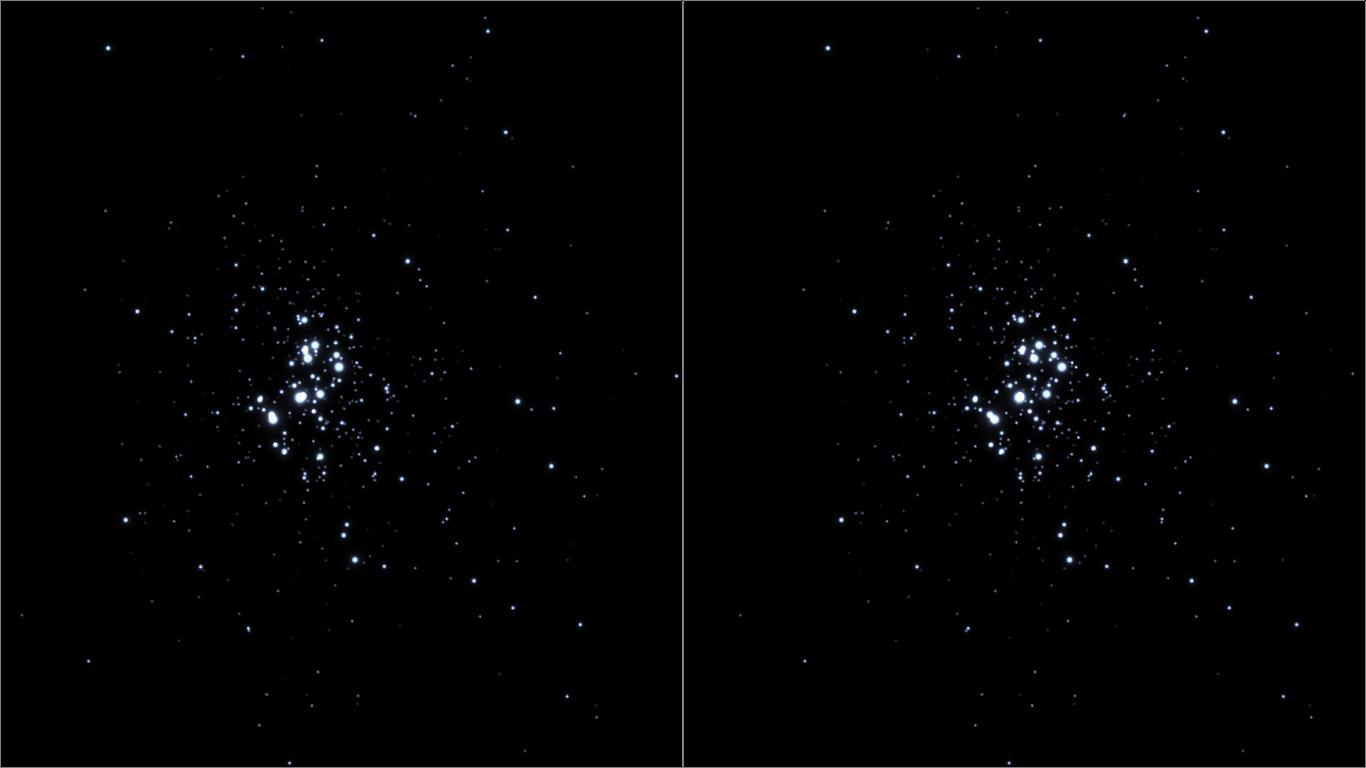}
\caption{Stereoscopic image of the Pleiades. The simulated separation between the eyes is 3 light years. Place the image around 40 cm away and cross your eyes until both images fuse into a three dimensional view in between. (Done with Celestia.)}
\label{stereo}
\end{figure}

Such a precise determination of the positions allows us to plot the cluster in 3D. We show this in Fig.~\ref{3dplot}, where each star is a sphere according to its magnitude. The brightest stars, visible to the naked eye in the famous cluster, are shown in red. We can see that the swarm is rather spherical, and that the brightest stars are aligned forming a column pointing towards us (to the bottom of the plot). In Fig.~\ref{sideways} we show a realistic visualization of the Pleiades seen ``from the side,'' an unusual view from an imaginary starship passing by. A short video flying around the Pleiades can be found at \url{http://youtu.be/nXS-13gCxh8}. Finally, in Fig.~\ref{stereo} we have prepared a stereoscopic image to view with crossed eyes. We present them approximately as we see them from Earth, but as if we had a 3 light years separation between our eyes.

\section*{Appendix}

We provide here some calculation details. The data were dowloaded from the Gaia archive (\url{gea.esac.esa.int/archive}), searching for all the sources within a $5^{\circ}$ circle centered at $RA=56.75^{\circ}$, $DEC=24.12^{\circ}$. This produced $699~\!860$ sources.

To this set we applied the criterion of dispersion of the proper motion: $\sqrt{(p_\alpha - 20.5)^2+(p_\delta+45.5)^2} < 6$~mas/yr, (where $p_\alpha$ and  $p_\delta$ are the proper motions in right ascension and declination respectively), finding 1876 stars. This criterion is probably very restrictive, since we had to manually add Merope, for which it gives $6.76$~mas/yr. There are probably more members in the cluster, but this is the criterion used in \citet{dr1} and it was used exactly as such.

We finally selected the 1595 stars with parallax betwen 5 and 9.5. 

The weighted mean was calculated using weights $w_i=1/\sigma_i^2$, where $\sigma_i$ is each parallax error according to Gaia DR2. This choice provides the maximum likelihood of the estimate of the mean, assuming that each measurement is a random variable with gaussian distribution of standard deviation $\sigma_i$ \citep{bevington}. In this case, the error of the estimate is $(\sum w_i)^{-1/2}$. This assumes that the errors of individual measurements are independent, which is not the case for Gaia data so close one to the other. \citet{dr1} gives the recommendation of adding 0.3~mas as a systematic error, which we haven't done here for our initial estimate. The error in distance reported above corresponds to the propagation of the relative error when converting parallax to distance. 

A more sophisticated calculation would require an account of the systematic error, but \citet{luri2018} explicitly says that ``unfortunately, there is no simple recipe to account for the
systematic errors.'' For the specific case of stellar clusters, \citet{bailer-jones2017} sugests the use of a model of the distribution of stars in the cluster to infer the distance to its center. In addition, \citet{luri2018} recommends to perform a Bayesian analysis of the errors taking into account also magnitude and color. Such refinements lie beyond the scope of the present article and will eventually be presented elsewhere.

\section*{Acknowledgements}
This work has made use of data from the European Space Agency (ESA) mission Gaia (\url{www.cosmos.esa.int/gaia}), processed by the Gaia Data Processing and Analysis Consortium (DPAC, \url{www.cosmos.esa.int/web/gaia/dpac/consortium}). Funding for the DPAC has been provided by national institutions, in particular the institutions participating in the Gaia Multilateral Agreement.

\end{document}